\documentclass{pasj00}
\draft

\def\commenta{$^*$}
\def\commentb{$^\dagger$}
\def\commentc{$^\ddagger$}
\def\commentd{$^\S$}
\def\commente{$^\|$}
\def\commentf{$^\#$}

\DeclareAbbreviation\AAHam{Astron. Abh. Hamburg. Sternw.}
\DeclareAbbreviation\AARv{Astron. Astrophys. Rev.}
\DeclareAbbreviation\AAS{American Astron. Soc. Meeting Abstracts}
\DeclareAbbreviation\AcA{Acta Astron.}
\DeclareAbbreviation\actaa{Acta Astron.}
\DeclareAbbreviation\Afz{Astrofizika}
\DeclareAbbreviation\AGAb{Astronomische Gesellschaft Abstract Ser.}
\DeclareAbbreviation\an{Astron. Nachr.}
\DeclareAbbreviation\AnAp{Annales d'Astrophysique}
\DeclareAbbreviation\AnTok{Tokyo Astron. Obs. Annals, Sec. Ser.}
\DeclareAbbreviation\Ap{Astrophysics}
\DeclareAbbreviation\ARep{Astron. Rep.}
\DeclareAbbreviation\AstBu{Astrophys. Bull.}
\DeclareAbbreviation\ATel{Astron. Telegram}
\DeclareAbbreviation\ATsir{Astron. Tsirk.}
\DeclareAbbreviation\AcApS{Acta Astrophys. Sinica}
\DeclareAbbreviation\AstL{Astron. Lett.}
\DeclareAbbreviation\BaltA{Baltic Astron.}
\DeclareAbbreviation\BANS{Bull. of the Astron. Institutes of the Netherlands Suppl. Ser.}
\DeclareAbbreviation\BASI{Bull. Astron. Soc. India}
\DeclareAbbreviation\BeSN{Be Newslett.}
\DeclareAbbreviation\BHarO{Harvard Coll. Obs. Bull.}
\DeclareAbbreviation\CBET{Cent. Bur. Electron. Telegrams}
\DeclareAbbreviation\ChJAA{Chinese J. of Astron. and Astrophys.}
\DeclareAbbreviation\caa{Chinese J. of Astron. and Astrophys.}
\DeclareAbbreviation\CoAsi{Asiago Contr.}
\DeclareAbbreviation\CoSka{Contributions of the Astronomical Observatory Skalnat\'e Pleso}
\DeclareAbbreviation\GCN{GRB Coord. Netw. Circ.}
\DeclareAbbreviation\ErgAN{Erg. Astron. Nachr.}
\DeclareAbbreviation\ibvs{IBVS}
\DeclareAbbreviation\IEEEP{IEEE Proc.}
\DeclareAbbreviation\JAD{J. Astron. Data}
\DeclareAbbreviation\JAVSO{J. American Assoc. Variable Star Obs.}
\DeclareAbbreviation\JBAA{J. Br. Astron. Assoc.}
\DeclareAbbreviation\JPhCS{J. of Physics Conference Series}
\DeclareAbbreviation\JPSJ{J. Phys. Soc. Japan}
\DeclareAbbreviation\JSARA{J. of the Southeastern Assoc. for Research in Astron.}
\DeclareAbbreviation\LowOB{Lowell Obs. Bull.}
\DeclareAbbreviation\MitAG{Mitteil. der Astronom. Gesell. Hamburg}
\DeclareAbbreviation\MitVS{Mitteil. Ver\"{a}nderl. Sterne}
\DeclareAbbreviation\MmSAI{Mem. Soc. Astron. Ital.}
\DeclareAbbreviation\memsai{Mem. Soc. Astron. Ital.}
\DeclareAbbreviation\Msngr{Messenger}
\DeclareAbbreviation\NewA{New Astron.}
\DeclareAbbreviation\na{New Astron.}
\DeclareAbbreviation\NewAR{New Astron. Rev.}
\DeclareAbbreviation\nar{New Astron. Rev.}
\DeclareAbbreviation\NInfo{Nauchnye Informatsii}
\DeclareAbbreviation\OAP{Odessa Astron. Publ.}
\DeclareAbbreviation\Obs{Observatory}
\DeclareAbbreviation\OEJV{Open Eur. J. on Variable Stars}
\DeclareAbbreviation\PASA{Publ. Astron. Soc. Australia}
\DeclareAbbreviation\PASAu{Publ. Astron. Soc. Australia}
\DeclareAbbreviation\PAZh{Pis'ma AZh}
\DeclareAbbreviation\POBeo{Publ. de l'Observatoire Astronomique de Beograd}
\DeclareAbbreviation\PCCP{Phys. Chem. Chem. Phys.}
\DeclareAbbreviation\PhR{Phys. Rep.}
\DeclareAbbreviation\PVSS{Publ. Variable Stars Sect. R. Astron. Soc. New Zealand}
\DeclareAbbreviation\PZ{Perem. Zvezdy}
\DeclareAbbreviation\PZP{Perem. Zvezdy, Prilozh.}
\DeclareAbbreviation\QJRAS{QJRAS}
\DeclareAbbreviation\RA{Ricerche Astronomiche}
\DeclareAbbreviation\RMxAA{Rev. Mexicana Astron. Astrof.}
\DeclareAbbreviation\RvMA{Reviews of Modern Astron.}
\DeclareAbbreviation\SASS{Society for Astronom. Sciences Ann. Symp.}
\DeclareAbbreviation\Sci{Science}
\DeclareAbbreviation\SPIE{SPIE Proc.}
\DeclareAbbreviation\SvA{Soviet Astronomy}
\DeclareAbbreviation\SvAL{Soviet Astronomy Letters}
\DeclareAbbreviation\VeSon{Ver\"{o}ff. Sternw. Sonneberg}
\DeclareAbbreviation\VSOLJBul{VSOLJ Variable Star Bull.}
\DeclareAbbreviation\yCat{VizieR Online Data Catalog}
\DeclareAbbreviation\ZA{Z. Astrophys.}

\newcounter{author}
\def\authorcount#1#2{\refstepcounter{author}\label{#1}
\altaffiltext{\ref{#1}}{#2}}

\begin{document}
\SetRunningHead{C. Nakata et al.}{OT J075418.7+381225 and OT
J230425.8+062546: Promising Candidates for the Period Bouncer}

\Received{201X/XX/XX}
\Accepted{201X/XX/XX}

\title{OT J075418.7+381225 and OT J230425.8+062546: Promising Candidates for 
the Period Bouncer}

\author{Chikako~\textsc{Nakata},\altaffilmark{\ref{affil:Kyoto}*}
Taichi~\textsc{Kato},\altaffilmark{\ref{affil:Kyoto}}
Daisaku\textsc{Nogami},\altaffilmark{\ref{affil:Kyoto}}
Elena~\textsc{Pavlenko},\altaffilmark{\ref{affil:CrAO}}$^,$\altaffilmark{\ref{affil:Kyoto}}
Tomohito~\textsc{Ohshima},\altaffilmark{\ref{affil:Kyoto}}
Enrique~de~\textsc{Miguel},\altaffilmark{\ref{affil:Miguel}}$^,$\altaffilmark{\ref{affil:Miguel2}}
William~\textsc{Stein},\altaffilmark{\ref{affil:Stein}}
Kazuhiko~\textsc{Siokawa},\altaffilmark{\ref{affil:Siz}}
Etienne~\textsc{Morelle},\altaffilmark{\ref{affil:Morelle}}
Hiroshi~\textsc{Itoh},\altaffilmark{\ref{affil:Ioh}}
Pavol~A.~\textsc{Dubovsky},\altaffilmark{\ref{affil:Dubovsky}}
Igor~\textsc{Kudzej},\altaffilmark{\ref{affil:Dubovsky}}
Hiroyuki~\textsc{Maehara},\altaffilmark{\ref{affil:Kiso}}
Arne~\textsc{Henden},\altaffilmark{\ref{affil:AAVSO}}
William~N.~\textsc{Goff},\altaffilmark{\ref{affil:Goff}}
Shawn~\textsc{Dvorak},\altaffilmark{\ref{affil:Dvorak}}
Oksana~\textsc{Antonyuk},\altaffilmark{\ref{affil:CrAO}}
Eddy~\textsc{Muyllaert},\altaffilmark{\ref{affil:VVSBelgium}}
}

\authorcount{affil:Kyoto}{
Department of Astronomy, Kyoto University, Kyoto 606-8502}
\email{$^*$nakata@kusastro.kyoto-u.ac.jp}

\authorcount{affil:CrAO}{
Crimean Astrophysical Observatory, 98409, Nauchny, Crimea, Ukraine}

\authorcount{affil:Miguel}{
Departamento de F\'isica Aplicada, Facultad de Ciencias
Experimentales, Universidad de Huelva,
21071 Huelva, Spain}

\authorcount{affil:Miguel2}{
Center for Backyard Astrophysics, Observatorio del CIECEM,
Parque Dunar, Matalasca\~nas, 21760 Almonte, Huelva, Spain}

\authorcount{affil:Stein}{
6025 Calle Paraiso, Las Cruces, New Mexico 88012, USA}

\authorcount{affil:Morelle}{
9 rue Vasco de GAMA, 59553 Lauwin Planque, France}

\authorcount{affil:Siz}{
     Moriyama 810, Komoro, Nagano 384-0085}

\authorcount{affil:Ioh}{
VSOLJ, 1001-105 Nishiterakata, Hachioji, Tokyo 192-0153}

\authorcount{affil:Dubovsky}{
Vihorlat Observatory, Mierova 4, Humenne, Slovakia}

\authorcount{affil:Kiso}{Kiso Observatory, Institute of Astronomy,
School of Science, The University of Tokyo,
10762-30, Mitake, Kiso-machi, Kiso-gun, Nagano 397-0101, Japan}

\authorcount{affil:AAVSO}{
     American Association of Variable Star Observers, 49 Bay State Rd.,
     Cambridge, MA 02138, USA}

\authorcount{affil:Goff}{
13508 Monitor Ln., Sutter Creek, California 95685, USA}

\authorcount{affil:Dvorak}{
Rolling Hills Observatory, 1643 Nightfall Drive,
Clermont, Florida 34711, USA}

\authorcount{affil:VVSBelgium}{
Vereniging Voor Sterrenkunde (VVS), Moffelstraat 13 3370
Boutersem, Belgium}



\maketitle

\begin{abstract}
We report on photometric observations of two dwarf novae,
OT J075418.7+381225 and OT J230425.8+062546,
which underwent superoutburst in 2013
(OT J075418) and in 2011 (OT J230425).
Their mean period of the superhump was 0.0722403(26) d (OT J074518)
and 0.067317(35) d (OT J230425).
These objects showed a very long growing stage of the superhump (stage A)
and a large period decrease in stage A-B transition.
The long stage A suggests slow evolution of the
superhump due to very small mass ratios of these objects.
The decline rates during the plateau phase in the superoutburst of these
objects were lower than those of SU UMa-type
DNe with a similar superhump period.
These properties were similar to those of SSS J122221.7$-$311523,
the best candidate for the period bouncer.
Therefore, these two DNe are regarded as good candidates for
the period bouncer.
We estimated the number density of period bouncers roughly 
from our observations in the recent five years. 
There is a possibility that these WZ Sge-type dwarf novae 
with unusual outburst properties
can account for the missing population of the period bouncer 
expected from the evolutionary scenario.

\end{abstract}

\section{Introduction}\label{sec:intro}

Cataclysmic variables (CVs) are binary star systems composed
of a white dwarf (primary) and a secondary which is typically
a late-type main sequence star. The secondary fills its Roche
lobe and matter falls down toward the primary spilling over
from the inner lagrangian point ($L_1$ point).

Dwarf novae (DNe) are one of subtypes of CVs. DNe undergo recurring
outbursts. The outburst lasts for an order of days to weeks, 
during which their brightness increases by 2 to 5 mag.
The outburst results from a release of gravitational
energy which is caused by a sudden increase 
of the mass accretion rate 
by the thermal instability in the disk.

SU UMa-type dwarf novae are a subclass of DNe. They have
relatively short orbital periods (1--2 hrs, near to 
the period minimum) and
occasional ``superoutbursts" that are brighter and have longer durations
than the normal outbursts.
The superhumps are believed to result from the tidal instability
that is triggered when the disk radius reaches the critical radius
for the $3:1$ resonance \citep{osa89suuma}.
WZ Sge-type DNe are a subgroup of SU UMa-type DNe.
They have particularly short orbital
periods and show infrequent large-amplitude superoutbursts
[for general properties of WZ Sge-type DNe, see e.g. \citet{bai79wzsge};
\citet{dow90wxcet}; \citet{kat01hvvir}].
During the superoutbursts, superhumps, periodic light variations
whose period is a few percent shorter than the orbital period,
are seen. The superhump periods vary through a course of three 
stages: the first is the stage A with a longer superhump period, 
the middle is the stage B with a systematically varying period, 
and the final is the stage C
with a shorter superhump period \citep{Pdot}.

According to the standard evolutionary theory of CVs, the mass 
transfer from the secondary starts when the secondary 
fills its Roche lobe. The orbital period $P_{\rm orb}$ is 
longer when a CV is formed and the system evolves with 
$P_{\rm orb}$ becoming shorter. Once its $P_{\rm orb}$ reaches 
the period minimum, the secondary becomes oversized 
for its mass as a result
of deviation from thermal equilibrium or becomes 
a brown dwarf which
cannot maintain in hydrogen burning. After this point, the system
evolves toward longer period and it is usually called ``period bouncer"
[see e.g. \citet{kni11CVdonor} and references there in, for 
standard evolutionary theory of CVs].

The study about the period bouncers ought to play a vital role in 
resolving the problems about the terminal evolution of CVs, 
since it is said that \citet{kol93CVpopulation} estimated that 
70\% of CVs should have
passed the period bounce.
The candidates for the period bouncer, however, 
have hardly been discovered. 
One of the reasons is that
CVs become much fainter as they approach the period minimum
\citep{pat11CVdistance}.
\citet{lit06j1035} also gave a great impact on the problem about 
the missing population in the CVs.
They confirmed the secondary in the eclipsing short-period CV
SDSS 103533.03+055158.4 was a brown dwarf, which suggests that
the system is the period bouncer.
Recently, \citet{lit08eclCV} discovered more three systems which 
have a brown dwarf 
secondary with high-speed three-color photometry.
In photometric research in the period bouncers, until recently, 
WZ Sge-type DNe with multiple rebrightenings such as EG Cnc have 
been considered to be a good candidate for the period bouncer
\citep{pat98egcnc}.
Recently,
\citet{kat13qfromstageA} succeeded in interpreting the variation of
the superhump period around the stage A and developed 
a new dynamical method to estimate
the binary mass ratios ($q\equiv M_2/M_1$) only from the 
stage A superhump observations and the orbital period.
Using this new method,  
it become evident 
that many of WZ Sge-type DNe with multiple rebrightenings do 
not likely have low mass ratios as
estimated in EG Cnc \citep{nak13j2112j2037}.
After this suggestion, a new candidate for the period bouncer 
was discovered; 
\citet{kat13j1222} reported SSS J122221.7$-$311523 (hereafter SSS J122221), 
a transient discovered by Catalina Real-time Transient Survey 
(CRTS, \cite{CRTS}) Siding Spring Survey (SSS), 
had a very small mass ratio $q=0.045$ and long orbital period 
[possible period of 0.075879(1) d].
They also revealed a characteristic property of 
SSS J122221  
that stage A superhumps lasted for long time.

In this paper, we present two DNe which are similar in some properties
to SSS J122221.
OT J075418.7+381225 (hereafter OT J075418) was
detected by CRTS 
as CSS 130131 on 2013 January 31.
The quiescent counterpart was $g$=22.8 mag
SDSS J075418.72+381225.2.
The observed superhumps with a period of ~0.07 d were
suggestive of an SU UMa-type dwarf nova (vsnet-alert 15355).
OT J230425.8+062546 (hereafter OT J230425) was originally
reported as a possible nova discovered by Hideo Nishimura
on 2010 December 29 at 13.7 mag \citep{nak11j2304cbet2616}. 
The quiescent counterpart was $g$=21.1 mag SDSS J230425.88+062545.6. 
After this,
it was suggested to be a dwarf nova on the basis of the color of
the SDSS counterpart (vsnet-alert 12548). Subsequent
observations detected the presence of superhumps with an amplitude
of 0.06 mag (A. Arai, vsnet-alert 12563).
Although observations and analysis of OT J230425 were already reported
as a summary form in \citet{Pdot3}, we present a new interpretation
of this object in this paper.

This paper is structured as follows. Section \ref{sec:obs} briefly
shows a log of observations and our analysis method. Sections
\ref{sec:resultj0754} and \ref{sec:resultj2304} deal with the
results of the observations of OT J075418 and OT J230425, respectively.
Section \ref{sec:discussion} discusses the results.

\section{Observation and Analysis}\label{sec:obs}

Tables \ref{tab:log} and \ref{tab:log_j2304} show the logs
of photometric observations.
All the observation times were written in barycentric Julian days (BJDs).
To correct zero-point of data differences between different observers,
we added a constant to each observer's data.

The phase dispersion minimization (PDM) method \citep{PDM}
was used in a period analysis.
In subtracting the global trend of the light curve,
we subtracted smoothed light curve obtained by locally-weighted polynomial
regression (LOWESS, \cite{LOWESS}) before making the PDM analysis.
The 1-$\sigma$ error of the best estimated period by the PDM
analysis was determined by the methods in \citet{fer89error}, \citet{Pdot2}.

A variety of bootstrapping was used to estimating the robustness of
the result of PDM.
We analyzed about 100 samples which randomly contain 50\%
of observations, and performed PDM analysis for these samples.
The result of bootstrap is displayed as a form of 90\% confidence
intervals in the resultant $\theta$ statistics.

\section{OT J075418.7+381225}\label{sec:resultj0754}

\subsection{Overall Light Curve}\label{sec:overallj0754}

\begin{figure}
\begin{center}
\FigureFile(88mm,110mm){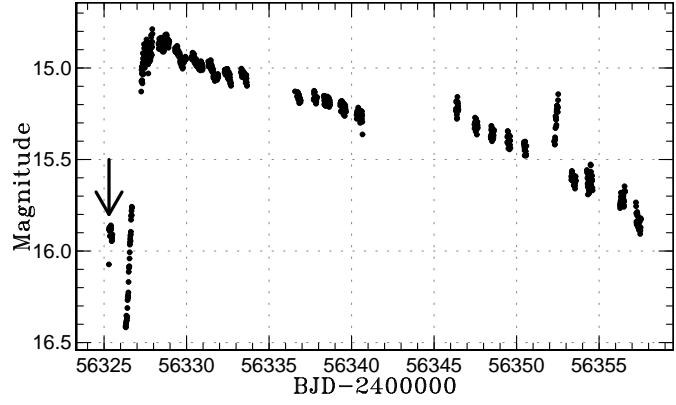}
\end{center}
\caption{Overall light curve of OT J075418. The data
were binned to 0.01 d. The arrow indicates the precursor.}
\label{fig:overall}
\end{figure}

Figure \ref{fig:overall} shows the overall light curve
of OT J075418.
After a precursor outburst (marked with an arrow in figure 
\ref{fig:overall}), the superoutburst started on
BJD 2456326. The early rise was well observed during BJD 
2456326--2456327. The superoutburst lasted with a 
slow decline for
at least 30 d.
In the middle part of the superoutburst (BJD 2456341--2456345),
there were no observations. On BJD 2456346, observations
were resumed and they showed a small rise of brightness.
On BJD 2456352, there was a rapid brightening.
This phenomenon was confirmed by
using different comparison stars.
It may have been an interesting phenomenon that we could not 
explain theoretically.
It, however, may have been an artifact, since the 
observing condition was very bad due to clouds and the moon.

\begin{figure*}
\begin{center}
\FigureFile(120mm,180mm){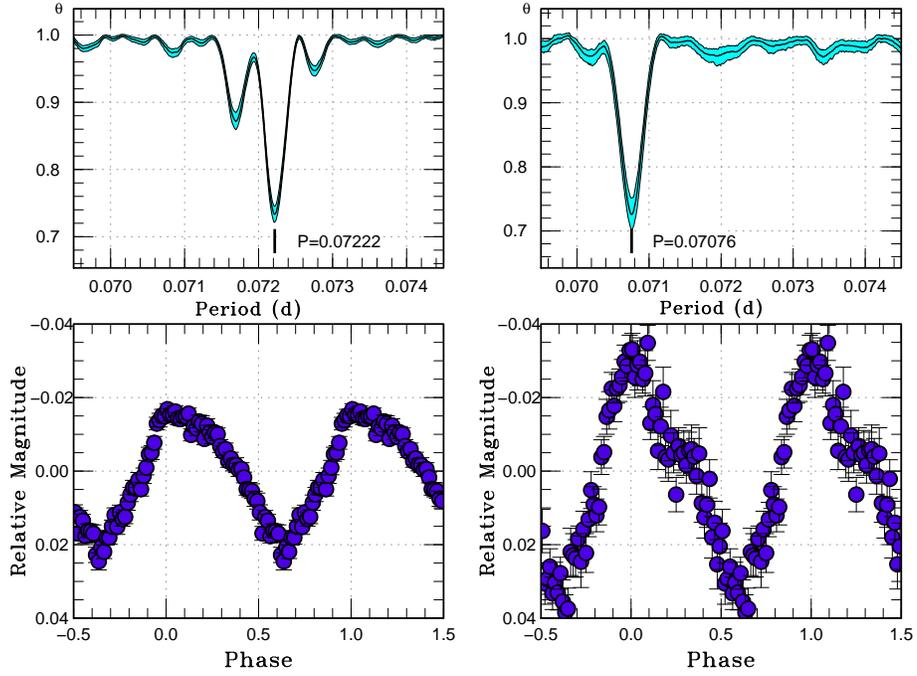}
\end{center}
\caption{Superhumps in OT J075418 (BJD 2456328--2456358).
(Left upper): $\theta$ diagram of our PDM analysis of 
stage A superhumps (BJD 2456328--2456341).
(Left lower): Phase-averaged profile of stage A superhumps.
(Right upper): $\theta$ diagram of our PDM analysis of 
stage B superhumps (BJD 2456345--2456355).
(Right lower): Phase-averaged profile of stage B superhumps.
}
\label{fig:shpdm}
\end{figure*}

\subsection{Superhumps}\label{sec:superhump}

Figure \ref{fig:shpdm} shows that
a period analysis using the Phase Dispersion Minimization 
(PDM) method \citep{PDM} indicated the presence of a period of
0.072218(3) d during stage A (BJD 2456328--2456341)
and
0.070758(6) d during stage B (BJD 2456345--2456355).
The mean profile of stage A and stage B superhumps
are also shown in lower part of figure \ref{fig:shpdm}.
The amplitude of the superhumps during stage B is larger than
that during stage A.

\begin{figure}
\begin{center}
\FigureFile(88mm,110mm){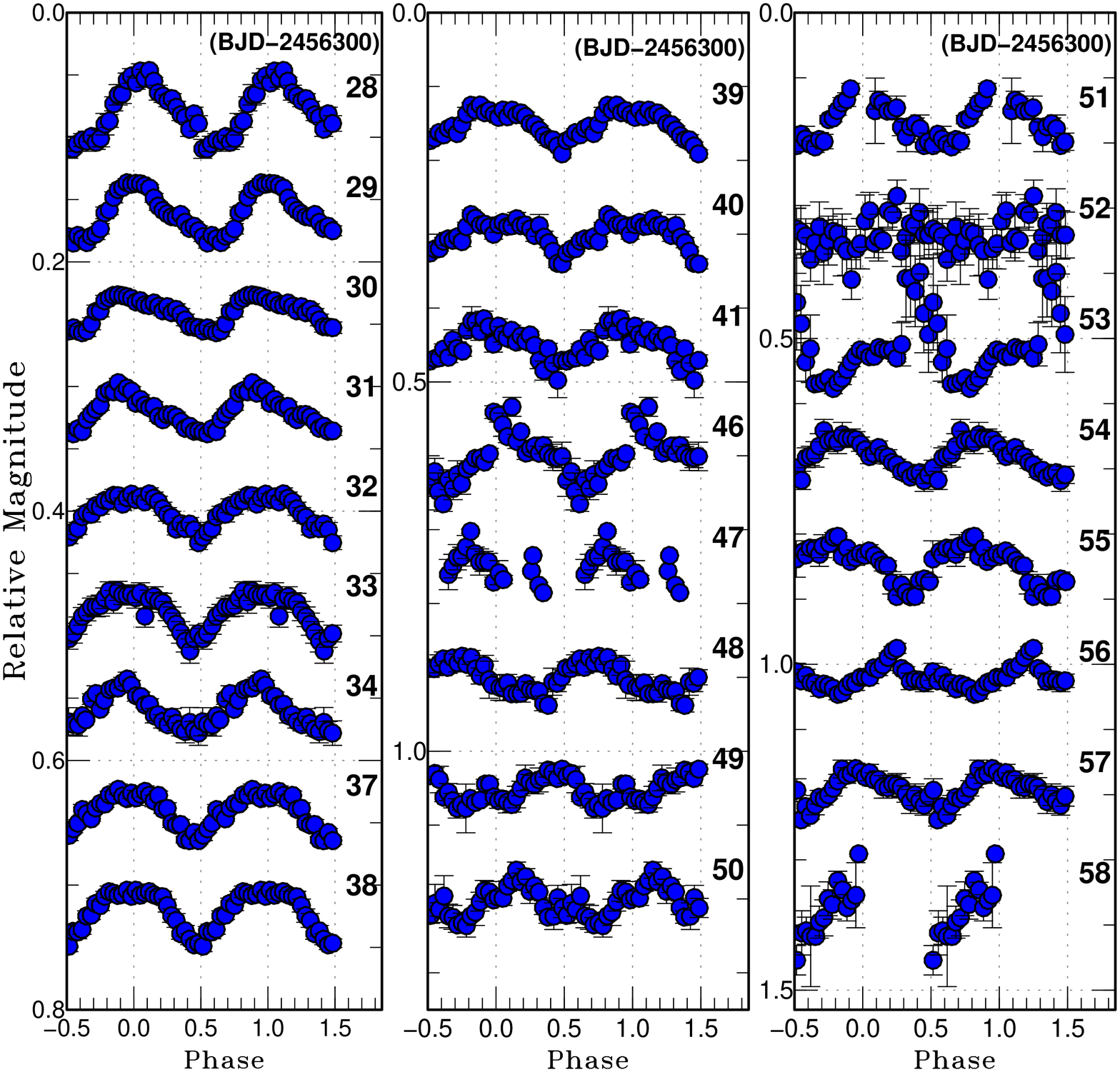}
\end{center}
\caption{Nightly variation of the profile of superhumps in OT J075418.}
\label{fig:pro_ordinary}
\end{figure}

Figure \ref{fig:pro_ordinary} shows the nightly variation of the
profile of superhumps. The amplitude of superhumps was 0.03--0.06
mag, smaller than in typical SU UMa-type DNe.

We determined the times of maxima of ordinary superhumps
as in the way described in \citet{Pdot}. The resultant times
are listed in table \ref{tab:shmax}.

\begin{figure}
\begin{center}
\FigureFile(88mm,110mm){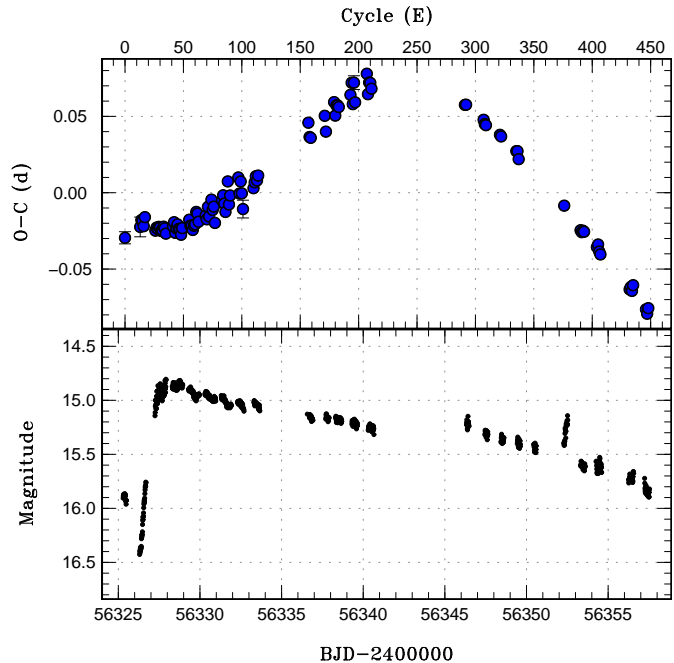}
\end{center}
\caption{(Upper:) The $O-C$ curve of OT J075418.
A ephemeris of
BJD 2456325.4414 + 0.0716368$E$ was used to draw this figure.
(Lower:) Overall light curve, the same as in figure \ref{fig:overall}.
The horizontal axis in units of BJD and the cycle number is common in 
both of upper and lower panels.}
\label{fig:ocplot}
\end{figure}

The $O-C$ curve of OT J075418 is shown in figure \ref{fig:ocplot}.
The very long stage A ($30\le E\le 220$) and stage B ($E\ge 280$) are seen.
Although the data when the stage A-B transition took place 
cannot be estimated precisely
because of lack of observations, it occurred between BJD 2456342 and 2456346.
In the stage A, superhumps with a mean period of
$P_{\rm sh}=$0.0722179(32) d 
and the time derivative of the superhump period 
$P_{\rm dot}$ ($=\dot{P}/P$) =$+3.6(0.7)\times 10^{-5}$ s/s
were recorded.
In the stage B, superhumps with a mean period of 
0.0707581(58) d and $P_{\rm dot}$ of
$-2.4(0.5)\times 10^{-5}$ s/s were recorded.

\subsection{Two-dimensional Lasso Analysis}\label{sec:lassoanalysis}

The least absolute shrinkage and selection operator (Lasso) method
was introduced by \citet{kat12perlasso}. This method has been
proven to be very effective in separating closely spaced periods and
has extended to two-dimentional power spectra (\cite{osa13v1504cygKepler};
\cite{kat13j1924}).

\begin{figure}
\begin{center}
\FigureFile(88mm,110mm){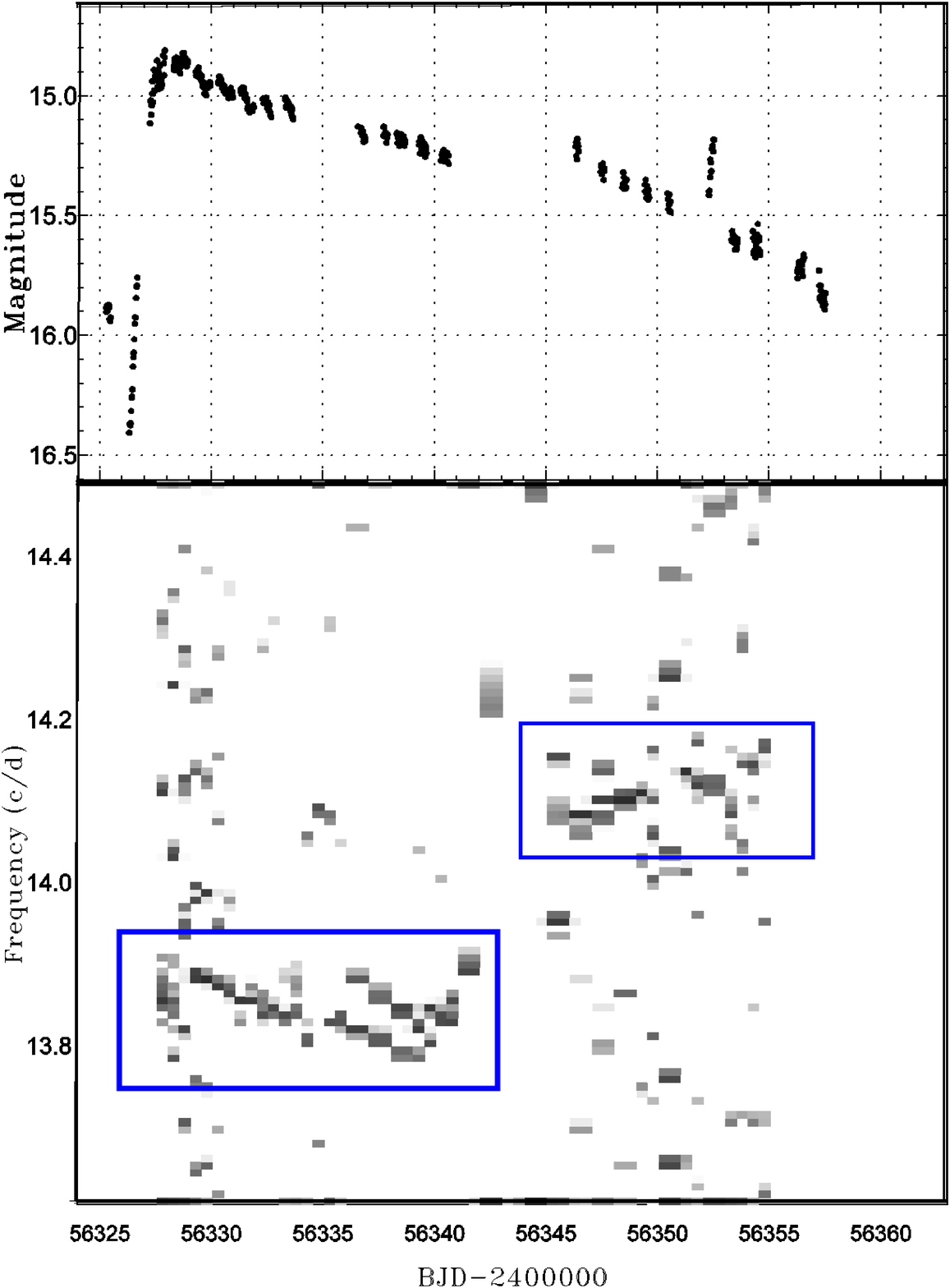}
\end{center}
\caption{Two-dimensional Lasso period analysis of OT J075418.
(Upper:) Overall light curve binned to 0.01 d, the same as 
figure \ref{fig:overall}. (Lower:) Result of two-dimensional
Lasso analysis (5 d window, 0.5 d shift and $\log \lambda = -8.5$). 
The appearance of the stage A and stage B frequency is boxed.}
\label{fig:lasso}
\end{figure}

A two-dimensional Lasso analysis of OT J075418 data is 
shown in figure
\ref{fig:lasso}.
A major change of frequency from $\sim 13.85$ $\rm{c/d}$ to
$\sim 14.1$ $\rm{c/d}$ can be
seen between BJD 2456341 and 2456345. It suggests that change coincided
with the timing when the stage A-B transition occurred.
During the stage A (BJD 2456328--2456341), the frequency
become lower. In contrast, it shows a tendency to become higher in 
the stage B (BJD 2456345--2456355).

\section{OT J230425.8+062546}\label{sec:resultj2304}
\subsection{Overall light curve}\label{sec:overallj2304}
\begin{figure}
\begin{center}
\FigureFile(88mm,110mm){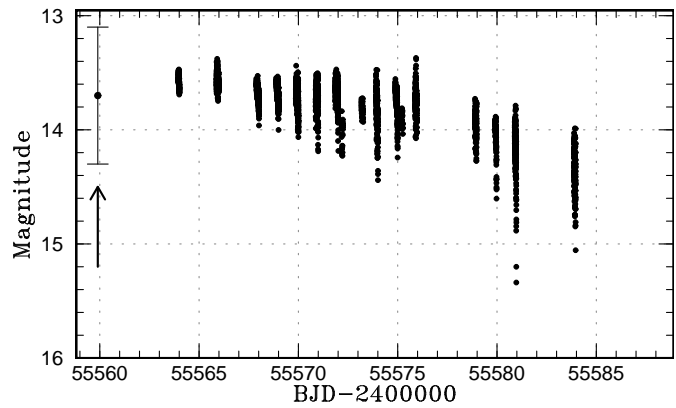}
\end{center}
\caption{Overall light curve of OT J230425. The data
were binned to 0.01 d. The arrow indicates the discovery of the 
superoutburst \citep{nak11j2304cbet2616}.}
\label{fig:overall_j2304}
\end{figure}

Figure \ref{fig:overall_j2304} shows the overall light curve
of OT J230425.
This object was discovered on December 29 in 2011 (BJD 2455559)
with the recorded possible maximum brightness of $V = 13.72$.
The early rise was missed. The superoutburst lasted for about
25 d. The light curve showed a slow decline until BJD 2455575.
After BJD 2455578, it declined faster.

\subsection{Superhumps}\label{sec:superhump_j2304}

During BJD 2455563--2455585, superhumps with amplitudes of
0.03--0.07 mag were present.
A period analysis using all the data indicated that 
the mean superhump period was
0.067317(35) d.
The PDM analysis of all superhumps was shown in \citet{Pdot3}.

\begin{figure*}
\begin{center}
\FigureFile(120mm,180mm){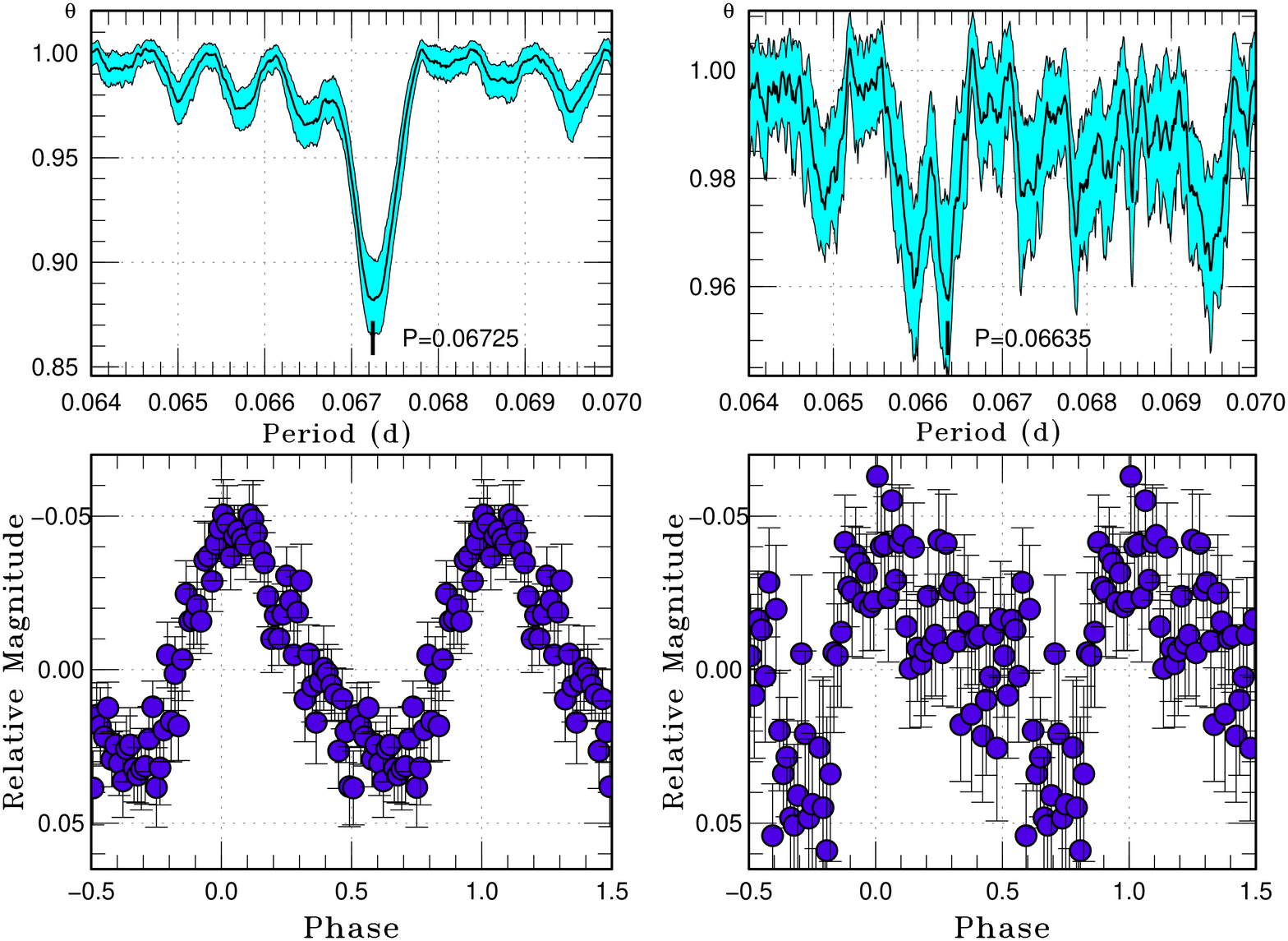}
\end{center}
\caption{Superhumps in OT J230425 (BJD 2455563--2455585).
(Left upper): $\theta$ diagram of our PDM analysis of stage A 
superhumps (BJD 2455563--2455572).
(Left lower): Phase-averaged profile of stage A superhumps.
(Right upper): $\theta$ diagram of our PDM analysis of stage B 
superhumps (BJD 2455571--2455585).
(Right lower): Phase-averaged profile of stage B superhumps.
}
\label{fig:shpdm_j2304}
\end{figure*}

\begin{figure}
\begin{center}
\FigureFile(88mm,110mm){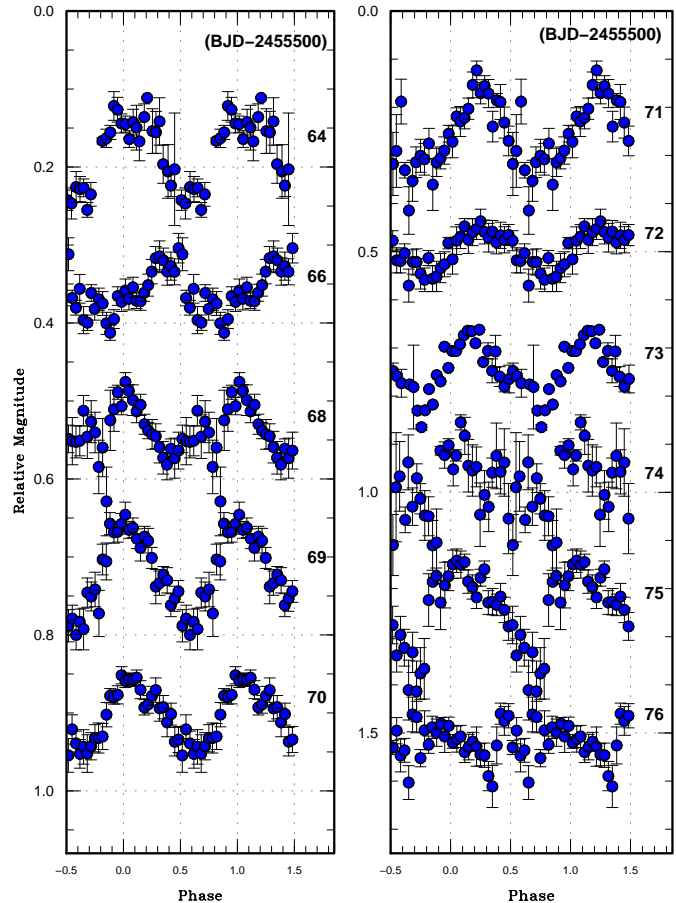}
\end{center}
\caption{Nightly variation of the profile of superhumps in OT J230425.}
\label{fig:pro_ordinary_j2304}
\end{figure}

A period analysis indicated a change of the period from
0.067245(17) d during stage A (BJD 2455563--2455572)
to 
0.066351(12) d during stage B (BJD 2455571--2455585) 
(figure \ref{fig:shpdm_j2304}).
The mean profile of stage A and stage B superhumps
are also shown in the lower panels of figure \ref{fig:shpdm_j2304}.
Figure \ref{fig:pro_ordinary_j2304} shows the nightly variation of the
profile of superhumps. The maximum amplitude of superhumps was
seen around BJD 2455571.

Displayed in top right panel of figure \ref{fig:shpdm_j2304}, 
there was a possible period which is shorter than 
the indicated period 0.066351(12) d. 
It was suggested that the period was a possible orbital 
period of 0.06589(1) d.
Assuming 0.06589(1) d to be the orbital period, 
the new method using stage A superhumps \citep{kat13qfromstageA} 
implies $q=0.053(1)$.
It suggests that OT J230425 is a good candidate for the period 
bouncer.

\begin{figure}
\begin{center}
\FigureFile(88mm,110mm){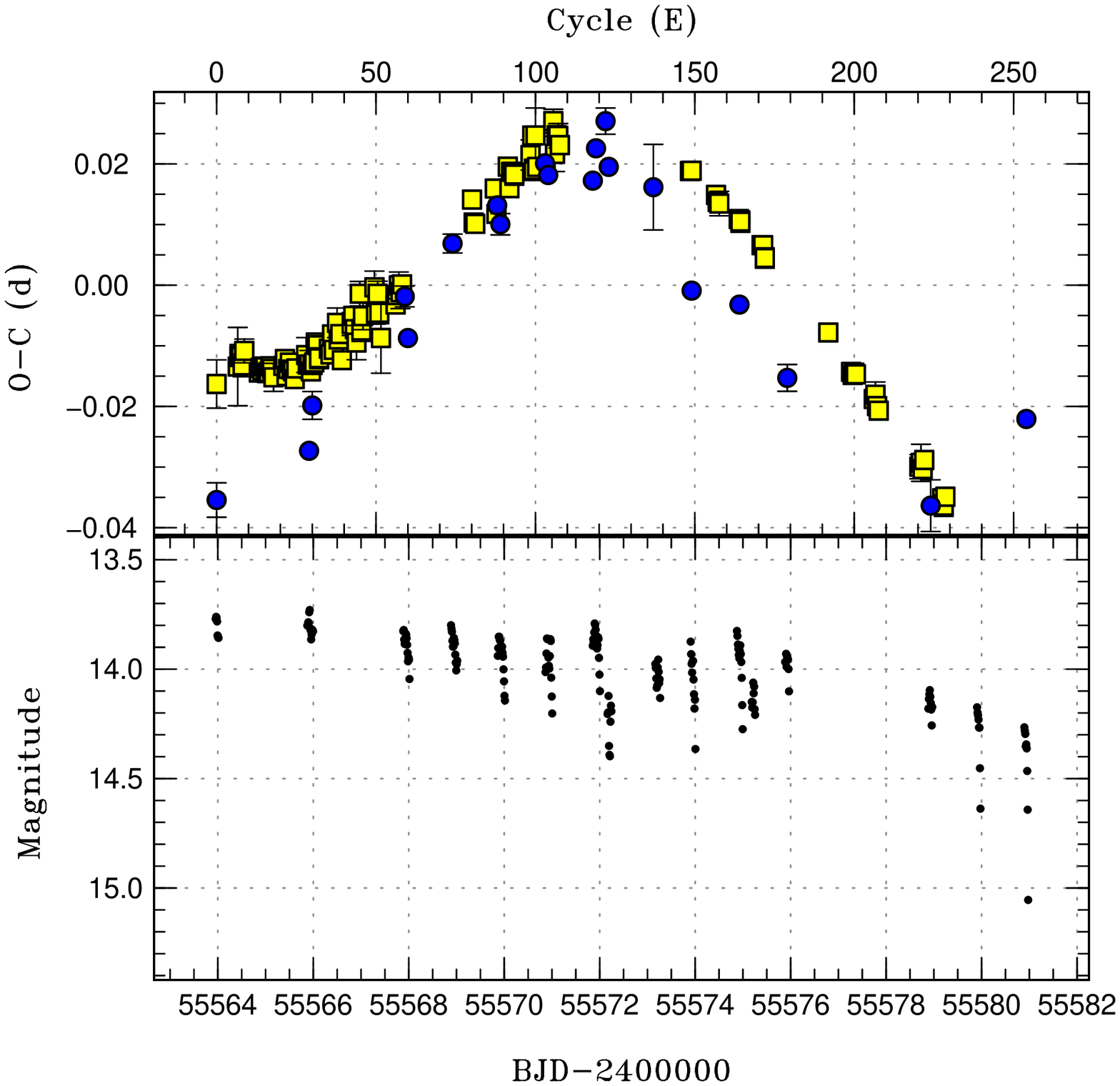}
\end{center}
\caption{(Upper:) $O-C$ curve of OT J230425, compared with 
that of OT J075418.
Filled circles and filled squares represent $O-C$
diagram of OT J230425 and that of OT J075418, respectively.
To fit two curves, the $O-C$ diagram of OT J075418 was shifted.
An ephemeris of
BJD 2455565.925 + 0.06695$E$ was used to draw this figure.
(Lower:) Light curve of OT J230425, the same as in figure 
\ref{fig:overall_j2304}.}
\label{fig:ocplot_j2304}
\end{figure}

Figure \ref{fig:ocplot_j2304} exhibits an $O-C$ curve of OT 
J230425 (filled circles), compared with $O-C$ curve of 
OT J075418 exhibited in figure \ref{fig:ocplot} (filled squares).
The resultant times of OT J230425 are listed in table \ref{tab:shmax_j2304}.
The $O-C$ curve of OT J230425 is very similar to that of OT J075418.
The very long stage A ($E\le 123$) and
the subsequent stage B ($E \ge 118$) are seen.
The stage A-B transition occurred around BJD 2455572.
The periods of superhumps in stage A and stage B were 0.067194(30) d
and 0.066281(63) d, respectively. The $P_{\rm dot}$ in stage B was
$-3.9(2.4)\times 10^{-5}$ s/s.

\section{Discussion}\label{sec:discussion}

\subsection{Decrease of Superhump Period between Stage A and
B}\label{sec:stageA-B}
The $O-C$ curves of OT J075418 and OT J230425 (figures \ref{fig:ocplot}
and \ref{fig:ocplot_j2304}) suggest a very long stage of increasing $O-C$ 
values (or a long period) and certain stage transition 
in the middle of the superoutbursts.
\citet{Pdot} argued that the superhump period usually decreases 
by 1.0--1.5\%
at the stage A-B transition and by $\sim 0.5\%$
at the stage B-C transition.
The fractional period decrease at the transition was $\sim2.0\%$ in OT
J075418 and $\sim1.4\%$ in OT J230425.
This large variation in frequency of OT J075418 can be clearly seen in
figure
\ref{fig:lasso}.
Since they were too large for the stage B-C transition, 
we regard this transition as the stage A-B transition.

The disk precession results mainly from the effects of 
direct axisymmetric
tidal potential from the secondary, secondarily from the 
gas pressure in the eccentric
mode and resonant wave stress \citep{lub92tilt}.
Although the tidal potential produces a net prograde precession, 
the gas pressure effect produces a retrograde contribution 
and decreases the precession rate.
\citet{mur00SHprecession} gave the hydrodynamical precession
$\omega$ in terms of the dynamic precession ($\omega_{\rm dyn}$) and
the pressure contribution to the precession ($\omega_{\rm pres}$):

\begin{equation}
\omega=\omega_{\rm dyn}+\omega_{\rm pres}.
\end{equation}

Note that $\omega_{\rm pres}$ is a negative value according to its 
retrograde contribution.
The ratio $\omega_{\rm pres}/\omega_{\rm orb}$ corresponds 
to the fractional
decrease of the superhump
period between stage A and B, where $\omega_{\rm orb}$ is the 
orbital frequency.
Therefore, it is possible that
the large decrease of the superhump period between stage A and B
indicates a large pressure contribution.

\subsection{Slow Evolution of Superhumps}\label{sec:longstageA}

The duration of the stage A reflects the growth time 
of the $3:1$ resonance.
As shown in subsections \ref{sec:superhump} and 
\ref{sec:superhump_j2304},
it took $\sim 190$ (OT J075418) and $\sim 120$ superhump cycles (OT
J230425), respectively 
to fully develop the $3:1$ resonance.
Considering absence of observation right after discovery of 
OT J230425, the growth time of the $3:1$ resonance may 
be even longer in OT J230425.
This long duration of the stage A suggests very small 
mass ratios $q$ of these
objects because the growth time of the $3:1$ resonance is expected to be
inversely proportional to $q^2$ \citep{lub91SHa}.
The duration of the stage A of these objects was 4--8 times longer
than those of typical SU UMa-type DNe having short 
orbital periods of $\sim 0.06$ d
and mass ratios of 0.10--0.15 \citep{Pdot}.
The mass ratios of these objects can be estimated
to be 2--3 times smaller, suggesting possible mass ratios $\sim 0.05$.
Despite the possible very small mass ratios, the orbital periods
of these objects, which are estimated
to be less than 1 \% shorter than their superhump periods, are
longer than that of typical short-period SU UMa 
($P_{\rm orb} \sim 0.06$ d).
This supports a hypothesis that these objects are 
candidates for the period bouncer.

\subsection{Slow Fading Rate}\label{sec:fadingrate}

During the superoutburst of SU UMa-type dwarf novae, an almost
exponential, slow decline phase exists, which is called 
the plateau phase.
\citet{osa89suuma} derived the time scale of this slow fading
as follows.
\begin{equation}\label{equ1}
t_{\rm d}\simeq 8.14 {\rm d} R_{\rm d,10}^{0.4} \alpha_{0.3}^{-0.7},
\end{equation}
where $R_{\rm d,10}$ is the disk radius in a unit of $10^{10}$ cm and
$\alpha_{0.3}=\alpha_{\rm hot}/0.3$, respectively ($\alpha_{\rm hot}$
represents the disk viscosity in the hot state).
\citet{Pdot5} suggested that $\alpha_{\rm hot}$ in candidates for 
the period bouncer that show slow fading rate is probably 
smaller than in higher-$q$ systems.
In addition to this,
the radius of the $3:1$ resonance can be formulated in terms of $q$:

\begin{equation}
r_{3:1}=3^{(-2/3)}(1+q)^{-1/3}.
\end{equation}

A small $q$, thus, produces a large radius of the $3:1$ resonance 
and a large disk radius.
But the contribution by a small $q$ is smaller than that by
a small $\alpha_{\rm hot}$, since equation \ref{equ1} shows 
the dependence of $t_{\rm d}$ to $q$ is larger than that of 
$\alpha$.

\begin{figure}
\begin{center}
\FigureFile(88mm,110mm){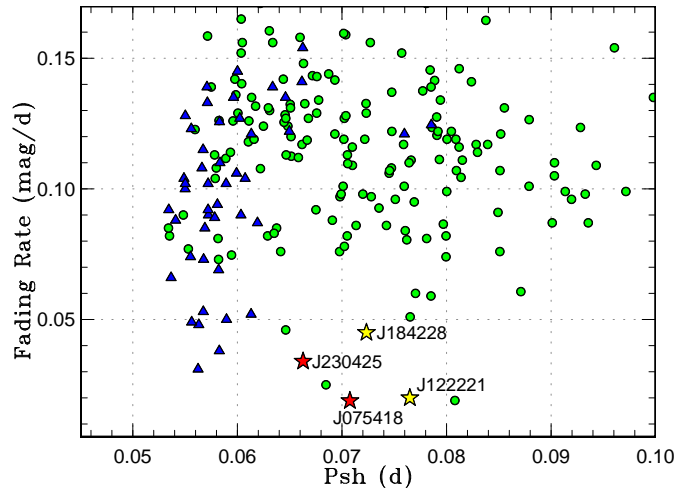}
\end{center}
\caption{Fading rate versus superhump period in the stage B.
The data are from \citet{Pdot5}. The filled circles and the
filled triangles represent SU UMa-type DNe and WZ Sge-type DNe,
respectively.
The filled stars represent the candidates for the period bouncer,
including OT J075418 and OT J230425.}
\label{fig:fadingrate}
\end{figure}

The fading rates of OT J075418 and OT J230425 were 0.0189(3)
mag $\rm{d^{-1}}$ and 0.0340(4) mag $\rm{d^{-1}}$, respectively.
Figure \ref{fig:fadingrate} shows the relation between the superhump
period in the stage B ($P_{\rm orb}$) and the fading rate of 
SU UMa-type DNe (filled circles), WZ Sge-type DNe 
(filled triangles) and
possible candidates for the period bouncer including OT J075418 and
OT J230425 (filled stars).
SSS J122221 (J122221 in figure \ref{fig:fadingrate}) was reported
as a perfect candidate for the period bouncer \citep{kat13j1222}.
\citet{kat13j1222} also suggests that
OT J184228.1+483742 (J184228 in figure \ref{fig:fadingrate}) showing
double superoutburst is a good candidate for the period bouncer.
According to figure \ref{fig:fadingrate}, the fading rates of 
OT J075418 and OT J230418 are
lower than those of SU UMa-type DNe with similar periods.
Furthermore, the location of these objects is close to that of
the good candidates for previously suggested the period bouncer.
This strengthens the interpretation that these objects are good candidates
for the period bouncer.

There are two  
objects near these candidates for the period bouncer 
in figure \ref{fig:fadingrate}.
One of them is BC Dor, which went through superoutburst in 2003, 
and another is PV Per detected its superoutburst in 2008.
Their last superoutbursts were reported in \citet{Pdot}.
Since they had relatively frequent outbursts, we considered 
them not to be candidates for the period bouncer.
They were inside a range of the period bouncer in error 
in figure \ref{fig:fadingrate} 
due to the data of poor quality.

\subsection{Absence of Early Superhumps}\label{sec:ab}
OT J184228, a likely candidate for the period bouncer, showed
double superoutburst consisting of the first one with early superhumps
and another one with ordinary superhumps \citep{kat13j1222}.
SSS J122221 also showed the similar pattern of the superoutburst.

OT J075418 and OT J230425, however, showed no early superhumps.
Early superhumps, as mentioned in section \ref{sec:intro},
arise when the disk radius reaches the 2:1 resonance
radius due to its low $q$.
And they cannot be detected in a system
with a low inclination (e.g. GW Lib reported by 
\cite{hir09gwlib}),
since the origin of early superhumps is
the emission of a
disk surface that has a non-axisymmetric vertical structure
(\cite{nog97alcom}; \cite{kat02wzsgeESH}).
The absence of the stage of early superhumps may indicate that 
the radius of the 2:1 resonance was not reached in these objects.
Although this generally does not support their low $q$,
we expect OT J075418 and OT J230425 to have low $q$ 
because of other strong evidence that we discussed above.
Furthermore, it has been reported that a system with 
corroborated low $q$ did not show early superhumps 
\citep{kat14j0902}.
It was considered that its 
superoutburst triggered by an inside-out (slowly rising) 
outburst made it impossible to establish the $2:1$ 
resonance. 

\subsection{Existence of a Precursor in OT J075418}
In OT J075418, a precursor preceding the superoutburst
was detected (BJD 2456325). This is rare in WZ Sge-type
DNe, which hardly show a normal outburst.

\citet{osa03DNoutburst} suggested that, in a superoutburst
with a precursor the disk radius does not reach the tidal 
truncation radius, which is the 
maximum radius larger than the $3:1$ resonance, while the 
disk radius reaches the tidal truncation radius in a 
superoutburst without a precursor.
It is thought that the system in a
precursor fades rapidly as in a normal outburst, since 
its accretion
disk does not reach the tidal truncation radius and the disk
allows the cooling wave to propagate inward.
During this fading, if the disk becomes sufficient eccentric,
the tidal dissipation from the secondary brings the disk
to the hot state. Then, it is observed as a superoutburst with a
precursor.
As mentioned in subsection \ref{sec:fadingrate}, however,
OT J075418 faded very slowly.
Similarly,
1RXS J053234+624755 has a very small mass ratio 0.074(19) and
showed a superoutburst with a precursor
(\cite{kap06j0532}; \cite{ima09j0532}).
Following \citet{ima09j0532}, we suggest that
the radius of the accretion disk of OT J075418 should be far beyond
the $3:1$ resonance, but not reach the tidal truncation radius,
if a mass ratio of a system is small enough to make a sufficient
space between the $3:1$ resonance and the tidal truncation radius.

As can be seen in the lightcurve, it took relatively a long time
from fading of a precursor outburst to appearance of a superoutburst.
It may be due to a long growth time of the $3:1$ resonance 
because of its small mass ratio.

\subsection{Number Density Problem of Period Bouncers}
Now, we know four candidates for the period bouncer (J075418, SSS J122221,
OT J230425 and OT J184228).
To investigate whether our observation is able to account for the 
missing population of period bouncers expected in the
evolutionary theory, we counted how many superoutbursts of 
SU UMa stars were observed.
In the recent five years, when the current project to observe SU UMa-type stars
was undertaken in the same way as now, we have observed about 291 superoutbursts
in 248 SU UMa-type stars (\cite{Pdot2}; \cite{Pdot3}; \cite{Pdot4};
\cite{Pdot5}).
The number of detected superoutbursts is unknown\footnote{There were 
systems that were detected its superoutburst but were not made 
time-series observations.}.

The recurrence time of the superoutbursts $T_{\rm s}$ is inversely
proportional to
mass-transfer rate $\dot{M}$ \citep{osa95wzsge}
and $\dot{M}$ is approximately proportional to $q^2$ in a 
short-period system 
evolving by gravitational radiation \citep{pat98evolution}. 
We can estimate the parent population of the period bouncers from 
the statistics of recorded outbursts.
Since many of SU UMa-type DNe have $T_{\rm s}\sim 1$ yr, we can assume 
many of them have been detected in a superoutburst in the last five years.
If the period bouncers have recurrence time of $T_{\rm s}(\rm{PB})$, 
the detection probability of period bouncers can be estimated 
to be $5/T_{\rm{s}}({\rm{PB}})\times f$, where $f$ stands for fraction of 
time covered by surveys.
If we conservatively assume $f\sim$ 0.1--0.5, 
we can thus estimate the ratio of parent populations of 
$N({\rm{PB}})/N$(ordinary SU UMa-type)$\sim 4/248\times 1/f \times 
T_{\rm{s}}({\rm{PB}})/5$.

There is large uncertainly in 
$T_{\rm s}({\rm{PB}})$.
We can, however, estimate $T_{\rm s}({\rm{PB}})>5$ yr, since 
these objects were hardly detected in outbursts in the past CRTS 
and other surveys.\footnote{Although typical intervals of 
observations in CRTS is 10 d, and there is a seasonal gap when 
the object is near the solar conjunction, we consider many 
(fraction $f$) 
of superoutburst should have been recorded 
since WZ Sge-type DNe usually show long-fading tails lasting 
several months.
No previous outbursts in four systems suggest that $T_{\rm s}$ 
for these systems are efficiently long.}
As to OT J230425, two outbursts have been detected by CRTS.
One outburst was in 2006 December and the other was in 2011 
January.
It suggests that the recurrence time of the outbursts of 
OT J230425 
is not as long as that of WZ Sge-type DNe.
Therefore, it is possible that some candidates for the period bouncer 
have higher mass-transfer rate than we expected and go through 
outbursts more frequent than WZ Sge-type DNe.
Three of the four candidates for the period bouncer are, however, 
WZ Sge-like stars.
We assume that majority of the candidates for the period bouncer
are WZ Sge-like stars and discuss the number density of period 
bouncers excluding the systems like OT J230425.
We regard $N({\rm{PB}})=3$ hereafter, excluding OT J230425. 

If we assume that the mass-transfer is purely driven by the 
gravitational wave radiation, 
$\dot{M}({\rm{PB}})\sim 10^{-2}\dot{M}$(ordinary SU UMa-type) 
and $T_{\rm s}({\rm{PB}})$ is expected to be $\sim10^2$ yr.
If we assume $T_{\rm{s}}({\rm{PB}})$ is $\sim10^1$ and $\sim10^2$ yr 
and conservatively assume $f\sim$ 0.1--0.5, 
we can obtain roughly $N({\rm{PB}})/N($SU UMa$)\sim$ 0.24--12.
Considering the possibility of the period bouncers like OT J230425, 
the population of the period bouncers can be estimated to be larger.
As mentioned in section \ref{sec:intro}, it was predicted that 
the majority of CVs ($\sim70\%$ by \cite{kol93CVpopulation}) 
have passed the period bounce.
On the contrary, few candidate for the period bouncer has 
been discovered by observations.
We call theoretically predicted population of the period bouncer 
``the missing population'' of the period bouncer. 
Although the true recurrence time of candidates for the period bouncer 
should be confirmed by future observations,
this ratio suggests a possibility that 
the period bouncers we have identified can 
account for the missing population of the period bouncers expected 
from the evolutionary scenario.
We thus identify these WZ Sge-type objects with unusual 
outburst properties are the good candidates for the 
hidden population of the terminal evolution of CVs.

Although we discussed photometric properties of the 
candidates for the period bouncer, \citet{gan09SDSSCVs} 
suggested their spectroscopic properties.
They argued that SDSS CVs in the 80--86 min period
spike showed spectra dominated by emission from the WD 
with no spectroscopic signature from the companion
star at optical wavelengths. 
These characteristics suggest that these systems have
very low accretion rates, and they are most likely 
DNe with extremely long recurrence time.
It takes long time to detect many candidates for 
the period bouncer in photometric observations on 
account of their long recurrence time. 
Spectroscopic studies of the newly identified four 
candidates are desired.

\section{Summary}\label{sec:summary}
We report on photometric observations of two dwarf novae,
OT J075418.7+381225 and OT J230425.8+062546,
which underwent superoutbursts in 2013
(OT J075418) and in 2011 (OT J230425).
The results of the analysis of our data 
are summarized in table \ref{tab:result}.

\begin{table}
\caption{The result of the analysis of OT J075418 and OT J230425.}
\label{tab:result}
\begin{center}
\begin{tabular}{ccc}
\hline
& OT J075418 & OT J230425 \\
\hline
mean period\commenta & 0.0722403(26) & 0.067317(35) \\
stage A\commentb & 0.072218(3) & 0.067245(17) \\
& (56328--56341) & (55563--55572) \\
stage B\commentc & 0.070758(6) & 0.066351(12) \\
& (56345--56355) & (55571--55585) \\
fading rate\commentd & 0.0189(3) & 0.0340(4) \\
\hline
\multicolumn{3}{l}{\commenta Mean period of the all superhumps. Unit d.} \\
\multicolumn{3}{l}{\commentb Stage A superhump period. Unit d.} \\
\multicolumn{3}{l}{The intervals used to determine the periods are given}\\
\multicolumn{3}{l}{in the parentheses. BJD$-$2400000.}\\
\multicolumn{3}{l}{\commentc Stage B superhump period. Unit d.} \\
\multicolumn{3}{l}{The intervals used to determine the periods are given}\\
\multicolumn{3}{l}{in the parentheses. BJD$-$2400000.}\\
\multicolumn{3}{l}{\commentd Unit mag $\rm{d^{-1}}$.}
\end{tabular}
\end{center}
\end{table}

In OT J075418 and OT J230425, some peculiar properties 
that were similar to
those of a good candidate for the period bouncer 
(SSS J122221.7$-$311523)
could be seen.
These two DNe are good candidates for
the period bouncer.
We then propose the general properties of candidates for 
the period bouncer as below:

\begin{itemize}
\item They show a very long growing stage of superhumps (stage A)
and a large period decrease of the stage A-B transition ($\sim1.5\%$).
The long stage A, which reflects the slow evolution of
the superhump, is due to the very small mass 
ratios of these objects.

\item The decline rates in the plateau phase in 
the superoutburst
of these objects are lower than those of SU UMa-type
DNe with the similar superhump period to these objects.
\end{itemize}

To investigate whether our observation is able to account for the 
missing population of the period bouncers expected in the
evolutionary theory, we counted how many SU UMa stars went through
superoutbursts.
In the recent five years, we have observed about 291 superoutbursts
in 248 SU UMa-type stars and four good candidates for the 
period bouncer 
have been suggested, including OT J075418 and OT J230425.
Three of four candidates were WZ Sge-like stars, and 
OT J230425 may have shorter recurrence time than the others.
We estimated the number density of the period bouncers, 
excluding the systems like OT J230425.  
Although there is large uncertainty in the recurrence time 
of the period bouncers, we assumed superoutbursts of 
the period bouncers 
were $10^1$--$10^2$ times more infrequent than those of 
ordinary SU UMa-type DNe, 
according to the theoretical prediction.
Under this assumption, 
we can obtain roughly $N({\rm{PB}})/N$(SU UMa)$\sim$ 0.24--12.
This ratio suggests a probability that the period bouncers 
we have identified can 
account for the missing population of the period bouncers expected 
from the evolutionary scenario.

\medskip
This work was supported by the Grant-in-Aid
“Initiative for High-Dimensional Data-Driven Science
through Deepening of Sparse Modeling” from the Ministry
of Education, Culture, Sports, Science and Technology
(MEXT) of Japan. 
We are grateful to many amateur observers for 
providing a lot of data used in this research.

\begin{table}
\caption{Log of observations of OT J075418.}\label{tab:log}
\begin{center}
\begin{tabular}{ccccccc}
\hline
Start\commenta & End\commenta & mag\commentb & error\commentc &
$N$\commentd & obs\commente & sys\commentf \\
\hline
25.2913 & 25.4934 & 15.947 & 0.003 & 229 & deM & C \\
26.2986 & 26.6882 & 16.191 & 0.010 & 489 & deM & C \\
27.2431 & 27.6752 & 15.120 & 0.003 & 433 & MEV & C \\
27.2980 & 27.6668 & 14.983 & 0.003 & 460 & deM & C \\
27.6840 & 27.9237 & 14.944 & 0.003 & 286 & GFB & C \\
28.2990 & 28.6527 & 14.923 & 0.001 & 441 & deM & C \\
28.5383 & 28.7812 & 14.988 & 0.002 & 290 & DKS & C \\
28.7016 & 28.9420 & 14.884 & 0.001 & 585 & SWI & V \\
29.2968 & 29.6605 & 14.969 & 0.001 & 514 & deM & C \\
29.6677 & 29.8488 & 15.019 & 0.002 & 200 & GFB & C \\
29.7057 & 29.9564 & 14.881 & 0.001 & 609 & SWI & V \\
30.2980 & 30.6371 & 15.005 & 0.001 & 500 & deM & C \\
30.7285 & 30.9638 & 14.909 & 0.001 & 572 & SWI & V \\
31.3028 & 31.6423 & 15.050 & 0.001 & 502 & deM & C \\
31.7016 & 31.9142 & 14.968 & 0.001 & 364 & SWI & V \\
32.3015 & 32.6886 & 15.105 & 0.001 & 608 & deM & C \\
32.3054 & 32.6595 & 15.138 & 0.001 & 355 & MEV & C \\
33.3011 & 33.6583 & 15.175 & 0.001 & 354 & MEV & C \\
36.5681 & 36.5681 & 15.180 & -- & 1 & MUY & C \\
36.6996 & 36.9058 & 1.169 & 0.001 & 353 & SWI & C \\
36.6996 & 36.9058 & 15.198 & 0.001 & 353 & SWI & V \\
37.6966 & 37.8978 & 1.166 & 0.001 & 345 & SWI & C \\
38.3036 & 38.6830 & 15.270 & 0.001 & 468 & CDZ & C \\
39.3021 & 39.5805 & 15.299 & 0.002 & 308 & CDZ & C \\
39.3564 & 39.6420 & 15.347 & 0.001 & 256 & MEV & C \\
\hline
\multicolumn{7}{l}{\commenta BJD$-$2456300.} \\
\multicolumn{7}{l}{\commentb Mean magnitude.} \\
\multicolumn{7}{l}{\commentc 1-$\sigma$ of the mean magnitude.} \\
\multicolumn{7}{l}{\commentd Number of observations.} \\
\multicolumn{7}{l}{\commente Observer's code. deM (E. de Miguel), MEV (E. Morelle),}\\
\multicolumn{7}{l}{ GFB (W. Goff), DKS (S. Dvorak), SWI (W. Stein),}\\
\multicolumn{7}{l}{ MUY (E. Muyllaert), CDZ (AAVSO data),}\\
\multicolumn{7}{l}{ DPV (P. Dubovsky)} \\
\multicolumn{7}{l}{\commentf Filter. ``C'' means no filter (clear).} \\
\end{tabular}
\end{center}
\end{table}

\addtocounter{table}{-1}
\begin{table}
\caption{Log of observations of OT J075418 (continued).}
\begin{center}
\begin{tabular}{ccccccc}
\hline
Start\commenta & End\commenta & mag\commentb & error\commentc &
$N$\commentd & obs\commente & sys\commentf \\
\hline
40.3004 & 40.6396 & 15.386 & 0.002 & 251 & MEV & C \\
40.3940 & 40.6534 & 15.346 & 0.002 & 293 & CDZ & C \\
46.3199 & 46.4654 & 15.350 & 0.003 & 148 & MEV & C \\
47.4372 & 47.6331 & 15.362 & 0.002 & 207 & deM & C \\
48.4312 & 48.6151 & 15.415 & 0.002 & 232 & deM & C \\
49.4269 & 49.6170 & 15.449 & 0.003 & 212 & deM & C \\
50.4413 & 50.5910 & 15.499 & 0.002 & 182 & deM & C \\
52.2781 & 52.5203 & 2.233 & 0.005 & 281 & DPV & C \\
53.3085 & 53.5875 & 15.658 & 0.002 & 354 & deM & C \\
54.2581 & 54.4964 & 2.554 & 0.003 & 300 & DPV & C \\
54.3650 & 54.6011 & 15.740 & 0.002 & 184 & MEV & C \\
56.2530 & 56.5597 & 2.655 & 0.002 & 386 & DPV & C \\
57.2314 & 57.5306 & 2.781 & 0.002 & 360 & DPV & C \\
\hline
\multicolumn{7}{l}{\commenta BJD$-$2456300.} \\
\multicolumn{7}{l}{\commentb Mean magnitude.} \\
\multicolumn{7}{l}{\commentc 1-$\sigma$ of the mean magnitude.} \\
\multicolumn{7}{l}{\commentd Number of observations.} \\
\multicolumn{7}{l}{\commente Observer's code. deM (E. de Miguel), MEV (E. Morelle),}\\
\multicolumn{7}{l}{ GFB (W. Goff), DKS (S. Dvorak), SWI (W. Stein),}\\
\multicolumn{7}{l}{ MUY (E. Muyllaert), CDZ (AAVSO data),}\\
\multicolumn{7}{l}{ DPV (P. Dubovsky)} \\
\multicolumn{7}{l}{\commentf Filter. ``C'' means no filter (clear).} \\
\end{tabular}
\end{center}
\end{table}

\begin{table}
\caption{Log of observations of OT J230425.}\label{tab:log_j2304}
\begin{center}
\begin{tabular}{ccccccc}
\hline
Start\commenta & End\commenta & mag\commentb & error\commentc &
$N$\commentd & obs\commente & sys\commentf \\
\hline
63.9532 & 64.0147 & 13.775 & 0.005 & 130 & Siz & C \\
65.8770 & 65.9577 & 14.149 & 0.004 & 146 & Mhh & C \\
65.8827 & 65.9319 & 13.683 & 0.005 & 107 & Ioh & C \\
65.8902 & 66.0023 & 13.783 & 0.002 & 224 & Siz & C \\
67.8853 & 68.0198 & 13.871 & 0.005 & 264 & Siz & C \\
68.8860 & 69.0162 & 13.876 & 0.004 & 259 & Siz & C \\
68.9449 & 69.0041 & 13.887 & 0.008 & 132 & Ioh & C \\
68.9769 & 68.9841 & 14.259 & 0.019 & 16 & Mhh & C \\
69.8702 & 70.0000 & 13.879 & 0.006 & 280 & Ioh & C \\
69.8802 & 69.9934 & 14.217 & 0.004 & 424 & Mhh & C \\
69.8892 & 70.0194 & 13.918 & 0.007 & 231 & Siz & C \\
70.8707 & 71.0046 & 13.908 & 0.007 & 268 & Ioh & C \\
71.8613 & 71.9994 & 13.851 & 0.005 & 292 & Ioh & C \\
71.8624 & 72.0124 & 13.848 & 0.006 & 299 & Siz & C \\
72.1643 & 72.2480 & 14.044 & 0.022 & 29 & CRI & C \\
73.1712 & 73.2678 & 13.813 & 0.008 & 47 & CRI & C \\
73.9053 & 73.9859 & 14.018 & 0.016 & 132 & Ioh & C \\
73.9302 & 74.0088 & 14.028 & 0.016 & 120 & Siz & C \\
74.8823 & 74.9998 & 13.951 & 0.009 & 236 & Siz & C \\
75.1832 & 75.2627 & 13.920 & 0.012 & 26 & CRI & C \\
75.8838 & 75.9353 & 14.272 & 0.009 & 182 & Mhh & C \\
75.9042 & 75.9720 & 13.962 & 0.006 & 134 & Siz & C \\
78.8920 & 78.9520 & 14.094 & 0.007 & 122 & Siz & C \\
78.9242 & 78.9722 & 14.508 & 0.008 & 182 & Mhh & C \\
79.9057 & 79.9786 & 14.275 & 0.015 & 120 & Siz & C \\
80.8940 & 80.9777 & 14.449 & 0.022 & 166 & Siz & C \\
80.9083 & 80.9452 & 14.588 & 0.009 & 136 & Mhh & C \\
83.8728 & 83.9645 & 14.549 & 0.016 & 158 & Ioh & C \\
86.9111 & 86.9111 & 17.401 & -- & 1 & Siz & C \\
\hline
\multicolumn{7}{l}{\commenta BJD$-$2455500.} \\
\multicolumn{7}{l}{\commentb Mean magnitude.} \\
\multicolumn{7}{l}{\commentc 1-$\sigma$ of the mean magnitude.} \\
\multicolumn{7}{l}{\commentd Number of observations.} \\
\multicolumn{7}{l}{\commente Observer's code. Siz (K. Shiokawa), Mhh (H. Maehara),}\\ 
\multicolumn{7}{l}{ Ioh (H. Itoh), CRI (Crimean Astrophys. Obs.)} \\
\multicolumn{7}{l}{\commentf Filter. ``C'' means no filter (clear).} \\
\end{tabular}
\end{center}
\end{table}

\begin{table}
\caption{Times of superhump maxima in OT J075418.}\label{tab:shmax}
\begin{center}
\begin{tabular}{ccccc}
\hline
$E$ & max\commenta & error & $O-C$\commentb & $N$\commentc \\
\hline
0 & 56325.4119 & 0.0040 & $-$0.0295 & 65 \\
13 & 56326.3502 & 0.0065 & $-$0.0224 & 72 \\
14 & 56326.4264 & 0.0018 & $-$0.0179 & 73 \\
15 & 56326.4975 & 0.0013 & $-$0.0184 & 71 \\
16 & 56326.5656 & 0.0019 & $-$0.0219 & 72 \\
17 & 56326.6432 & 0.0020 & $-$0.0160 & 74 \\
26 & 56327.2791 & 0.0005 & $-$0.0249 & 64 \\
27 & 56327.3527 & 0.0005 & $-$0.0229 & 136 \\
28 & 56327.4234 & 0.0004 & $-$0.0238 & 138 \\
29 & 56327.4965 & 0.0008 & $-$0.0224 & 125 \\
30 & 56327.5678 & 0.0004 & $-$0.0227 & 106 \\
31 & 56327.6370 & 0.0005 & $-$0.0251 & 139 \\
32 & 56327.7092 & 0.0004 & $-$0.0246 & 62 \\
33 & 56327.7832 & 0.0010 & $-$0.0222 & 64 \\
34 & 56327.8537 & 0.0007 & $-$0.0233 & 71 \\
35 & 56327.9219 & 0.0023 & $-$0.0268 & 44 \\
41 & 56328.3564 & 0.0007 & $-$0.0221 & 73 \\
42 & 56328.4308 & 0.0007 & $-$0.0193 & 73 \\
43 & 56328.4955 & 0.0009 & $-$0.0263 & 71 \\
44 & 56328.5700 & 0.0008 & $-$0.0234 & 138 \\
45 & 56328.6443 & 0.0011 & $-$0.0208 & 123 \\
46 & 56328.7130 & 0.0010 & $-$0.0236 & 157 \\
47 & 56328.7840 & 0.0005 & $-$0.0243 & 178 \\
48 & 56328.8524 & 0.0006 & $-$0.0275 & 141 \\
49 & 56328.9284 & 0.0010 & $-$0.0232 & 115 \\
55 & 56329.3637 & 0.0029 & $-$0.0177 & 87 \\
56 & 56329.4315 & 0.0012 & $-$0.0216 & 83 \\
57 & 56329.5031 & 0.0023 & $-$0.0215 & 82 \\
58 & 56329.5719 & 0.0013 & $-$0.0244 & 83 \\
59 & 56329.6463 & 0.0016 & $-$0.0216 & 61 \\
60 & 56329.7189 & 0.0007 & $-$0.0207 & 158 \\
\hline
\multicolumn{5}{l}{\commenta BJD$-$2400000.} \\
\multicolumn{5}{l}{\commentb $C$ = 2456325.4414 + 0.0716368$E$.} \\
\multicolumn{5}{l}{\commentc Number of points used to determine the
maximum.} \\
\end{tabular}
\end{center}
\end{table}

\addtocounter{table}{-1}
\begin{table}
\caption{Times of superhump maxima in OT J075418 (continued).}
\begin{center}
\begin{tabular}{ccccc}
\hline
$E$ & max\commenta & error & $O-C$\commentb & $N$\commentc \\
\hline
61 & 56329.7986 & 0.0014 & $-$0.0126 & 198 \\
62 & 56329.8692 & 0.0013 & $-$0.0137 & 157 \\
63 & 56329.9355 & 0.0017 & $-$0.0190 & 133 \\
69 & 56330.3692 & 0.0015 & $-$0.0151 & 85 \\
70 & 56330.4385 & 0.0011 & $-$0.0174 & 84 \\
71 & 56330.5183 & 0.0014 & $-$0.0092 & 86 \\
72 & 56330.5836 & 0.0010 & $-$0.0156 & 89 \\
74 & 56330.7380 & 0.0024 & $-$0.0045 & 78 \\
75 & 56330.8028 & 0.0009 & $-$0.0113 & 140 \\
76 & 56330.8767 & 0.0007 & $-$0.0091 & 141 \\
77 & 56330.9376 & 0.0011 & $-$0.0198 & 141 \\
83 & 56331.3822 & 0.0019 & $-$0.0050 & 80 \\
84 & 56331.4572 & 0.0011 & $-$0.0017 & 77 \\
85 & 56331.5234 & 0.0014 & $-$0.0071 & 95 \\
86 & 56331.5895 & 0.0028 & $-$0.0126 & 94 \\
88 & 56331.7528 & 0.0020 & 0.0074 & 99 \\
89 & 56331.8094 & 0.0018 & $-$0.0077 & 99 \\
90 & 56331.8869 & 0.0022 & $-$0.0018 & 99 \\
97 & 56332.4002 & 0.0027 & 0.0100 & 158 \\
98 & 56332.4612 & 0.0018 & $-$0.0006 & 145 \\
99 & 56332.5408 & 0.0021 & 0.0074 & 154 \\
100 & 56332.6046 & 0.0018 & $-$0.0004 & 152 \\
101 & 56332.6660 & 0.0058 & $-$0.0107 & 127 \\
110 & 56333.3244 & 0.0012 & 0.0030 & 45 \\
111 & 56333.3999 & 0.0008 & 0.0068 & 59 \\
112 & 56333.4755 & 0.0022 & 0.0108 & 61 \\
113 & 56333.5443 & 0.0027 & 0.0080 & 54 \\
114 & 56333.6192 & 0.0008 & 0.0112 & 61 \\
157 & 56336.7342 & 0.0008 & 0.0459 & 188 \\
158 & 56336.7965 & 0.0006 & 0.0365 & 198 \\
159 & 56336.8676 & 0.0013 & 0.0359 & 196 \\
171 & 56337.7416 & 0.0011 & 0.0504 & 99 \\
172 & 56337.8029 & 0.0012 & 0.0400 & 99 \\
\hline
\multicolumn{5}{l}{\commenta BJD$-$2400000.} \\
\multicolumn{5}{l}{\commentb $C$ = 2456325.4414 + 0.0716368$E$.} \\
\multicolumn{5}{l}{\commentc Number of points used to determine the
maximum.} \\
\end{tabular}
\end{center}
\end{table}

\addtocounter{table}{-1}
\begin{table}
\caption{Times of superhump maxima in OT J075418 (continued).}
\begin{center}
\begin{tabular}{ccccc}
\hline
$E$ & max\commenta & error & $O-C$\commentb & $N$\commentc \\
\hline
179 & 56338.3237 & 0.0015 & 0.0593 & 51 \\
180 & 56338.3865 & 0.0013 & 0.0505 & 71 \\
181 & 56338.4648 & 0.0013 & 0.0572 & 70 \\
182 & 56338.5365 & 0.0010 & 0.0572 & 71 \\
183 & 56338.6070 & 0.0018 & 0.0561 & 72 \\
193 & 56339.3314 & 0.0025 & 0.0641 & 74 \\
194 & 56339.4109 & 0.0011 & 0.0720 & 73 \\
195 & 56339.4686 & 0.0017 & 0.0580 & 131 \\
196 & 56339.5543 & 0.0045 & 0.0721 & 136 \\
197 & 56339.6130 & 0.0009 & 0.0592 & 69 \\
207 & 56340.3481 & 0.0020 & 0.0779 & 55 \\
208 & 56340.4063 & 0.0029 & 0.0645 & 91 \\
209 & 56340.4856 & 0.0013 & 0.0722 & 123 \\
210 & 56340.5570 & 0.0020 & 0.0719 & 89 \\
211 & 56340.6249 & 0.0013 & 0.0681 & 91 \\
291 & 56346.3452 & 0.0008 & 0.0575 & 47 \\
292 & 56346.4170 & 0.0007 & 0.0576 & 56 \\
307 & 56347.4816 & 0.0011 & 0.0478 & 50 \\
308 & 56347.5504 & 0.0015 & 0.0449 & 61 \\
309 & 56347.6214 & 0.0020 & 0.0442 & 49 \\
321 & 56348.4747 & 0.0011 & 0.0379 & 71 \\
322 & 56348.5453 & 0.0018 & 0.0368 & 70 \\
335 & 56349.4669 & 0.0012 & 0.0272 & 70 \\
336 & 56349.5385 & 0.0013 & 0.0272 & 46 \\
337 & 56349.6049 & 0.0017 & 0.0220 & 59 \\
376 & 56352.3683 & 0.0014 & $-$0.0085 & 70 \\
390 & 56353.3551 & 0.0011 & $-$0.0246 & 72 \\
391 & 56353.4255 & 0.0011 & $-$0.0258 & 70 \\
392 & 56353.4981 & 0.0012 & $-$0.0249 & 72 \\
393 & 56353.5691 & 0.0010 & $-$0.0256 & 63 \\
\hline
\multicolumn{5}{l}{\commenta BJD$-$2400000.} \\
\multicolumn{5}{l}{\commentb $C$ = 2456325.4414 + 0.0716368$E$.} \\
\multicolumn{5}{l}{\commentc Number of points used to determine the
maximum.} \\
\end{tabular}
\end{center}
\end{table}

\addtocounter{table}{-1}
\begin{table}
\caption{Times of superhump maxima in OT J075418 (continued).}
\begin{center}
\begin{tabular}{ccccc}
\hline
$E$ & max\commenta & error & $O-C$\commentb & $N$\commentc \\
\hline
404 & 56354.3470 & 0.0012 & $-$0.0357 & 74 \\
405 & 56354.4204 & 0.0021 & $-$0.0339 & 99 \\
406 & 56354.4874 & 0.0009 & $-$0.0385 & 105 \\
407 & 56354.5571 & 0.0010 & $-$0.0404 & 55 \\
432 & 56356.3251 & 0.0020 & $-$0.0633 & 71 \\
433 & 56356.3984 & 0.0031 & $-$0.0617 & 72 \\
434 & 56356.4674 & 0.0014 & $-$0.0643 & 70 \\
435 & 56356.5428 & 0.0012 & $-$0.0606 & 61 \\
446 & 56357.3150 & 0.0016 & $-$0.0764 & 67 \\
447 & 56357.3838 & 0.0018 & $-$0.0793 & 67 \\
448 & 56357.4591 & 0.0013 & $-$0.0756 & 70 \\
\hline
\multicolumn{5}{l}{\commenta BJD$-$2400000.} \\
\multicolumn{5}{l}{\commentb $C$ = 2456325.4414 + 0.0716368$E$.} \\
\multicolumn{5}{l}{\commentc Number of points used to determine the
maximum.} \\
\end{tabular}
\end{center}
\end{table}

\begin{table}
\caption{Times of superhump maxima in OT J230425.}\label{tab:shmax_j2304}
\begin{center}
\begin{tabular}{ccccc}
\hline
$E$ & max\commenta & error & $O-C$\commentb & $N$\commentc \\
\hline
0 & 55563.9791 & 0.0028 & $-$0.0354 & 108 \\
29 & 55565.9214 & 0.0011 & $-$0.0273 & 274 \\
30 & 55565.9956 & 0.0023 & $-$0.0198 & 64 \\
59 & 55567.9479 & 0.0017 & $-$0.0019 & 104 \\
60 & 55568.0077 & 0.0015 & $-$0.0087 & 70 \\
74 & 55568.9571 & 0.0016 & 0.0069 & 215 \\
88 & 55569.8972 & 0.0014 & 0.0132 & 347 \\
89 & 55569.9608 & 0.0018 & 0.0101 & 427 \\
103 & 55570.9046 & 0.0013 & 0.0201 & 116 \\
104 & 55570.9694 & 0.0011 & 0.0182 & 107 \\
118 & 55571.9022 & 0.0009 & 0.0172 & 217 \\
119 & 55571.9743 & 0.0014 & 0.0226 & 211 \\
122 & 55572.1788 & 0.0022 & 0.0271 & 15 \\
123 & 55572.2380 & 0.0014 & 0.0195 & 12 \\
137 & 55573.1684 & 0.0071 & 0.0162 & 16 \\
149 & 55573.9517 & 0.0014 & $-$0.0009 & 148 \\
164 & 55574.9499 & 0.0013 & $-$0.0032 & 107 \\
179 & 55575.9383 & 0.0022 & $-$0.0153 & 150 \\
224 & 55578.9187 & 0.0043 & $-$0.0364 & 237 \\
254 & 55580.9340 & 0.0014 & $-$0.0221 & 192 \\
\hline
\multicolumn{5}{l}{\commenta BJD$-$2400000.} \\
\multicolumn{5}{l}{\commentb $C$ = 2455565.925 + 0.06695$E$.} \\
\multicolumn{5}{l}{\commentc Number of points used to determine the
maximum.} \\
\end{tabular}
\end{center}
\end{table}

\newcommand{\noop}[1]{}

\end{document}